# Infrared Spectra and Optical Constants of Astronomical Ices: II. Ethane and Ethylene

R. L. Hudson, P. A. Gerakines, and M. H. Moore

Astrochemistry Laboratory
NASA Goddard Space Flight Center
Greenbelt, MD  20771
http://science.gsfc.nasa.gov/691/cosmicice

## Abstract

Infrared spectroscopic observations have established the presence of hydrocarbon ices on Pluto and other TNOs, but the abundances of such molecules cannot be deduced without accurate optical constants (*n*, *k*) and reference spectra.  In this paper we present our recent measurements of near- and mid-infrared optical constants for ethane ($C_2H_6$) and ethylene ($C_2H_4$) in multiple ice phases and at multiple temperatures.  As in our recent work on acetylene ($C_2H_2$), we also report new measurements of the index of refraction of each ice at 670 nm.  Comparisons are made to earlier work where possible, and electronic versions of our new results are made available.

## 1. Introduction

Solid methane is known to exist as a major component of the surface ices of both Pluto and Triton (e.g., Cruikshank et al., 1976; Cruikshank and Apt, 1984), and larger hydrocarbons are the subjects of present-day observational searches (e.g., DeMeo et al., 2010).  As New Horizons approaches the Pluto system, our research group is carrying out new IR optical-constants measurements of hydrocarbon and other ices with an emphasis on temperatures below ~70 K.  Our goal is to add to the relatively meager literature on this subject and to provide electronic versions of state-of-the-art data.  Additional applications are envisioned to areas of study and astronomical objects other than Pluto, such as Titan ices, icy moons, interstellar ices, and laboratory studies of ice chemistry by both reflection and transmission IR measurements.

In a recent paper we presented new measurements of the near- and mid-infrared (IR) optical constants, *n* and *k*, of amorphous and crystalline phases of solid acetylene ($C_2H_2$) at ~10 - 70 K (Hudson et al., 2014).  Such results are needed in order to model spectra of cold Solar System bodies suspected or known to contain hydrocarbons (e.g., Sasaki et al., 2005; Brown et al., 2007; Merlin et al., 2010; DeMeo et al., 2010) and, in a different context, to estimate acetylene abundances in interstellar clouds (e.g., Knez et al., 2012). Since such optical constants cannot yet be computed accurately *ab initio*, laboratory experiments are required for their determination.

The next-heavier common $C_2$ hydrocarbons beyond acetylene are ethane ($C_2H_6$) and ethylene ($C_2H_4$, sometimes called ethene), and they are the subject of the present paper. In terms of Pluto and Triton, photochemical models of their atmospheres predict both of these molecules as abundant precipitating products (e.g., Krasnopolsky and Cruikshank, 1995, 1999), and their observed reflectance spectra are well fit by laboratory spectra when pure, solid ethane is included as a component (e.g., Cruikshank et al., 2006). Our earlier brief publication on ethane ices (Hudson et al., 2009) described previous spectroscopic studies, and so only a summary is needed here. Ethane forms at least five solid phases, but since two exist only in a narrow (< 1 K) temperature region near 90 K, too warm for TNOs, they are ignored in the present paper. The remaining three phases of ethane are termed amorphous, crystalline, and metastable (or metastable crystalline). Amorphous ethane is formed by slow condensation of $C_2H_6$ gas below ~30 K, while condensation at ~30 - 55 K gives the metastable form (Wisnosky et al., 1983), and deposition at ~60 - 70 K in a vacuum gives the crystalline form. We note that crystalline $C_2H_6$ also has been termed phase II ethane (Pearl et al., 1991; Quirico and Schmitt, 1997), phase III ethane (Schutte et al., 1987), and ethane's α phase (Konstantinov et al., 2006). Warming either the amorphous or the metastable phases always produces crystalline $C_2H_6$ near 40 K and 65 K, respectively, and such phase changes are irreversible. The metastable phase cannot be reached by simply warming the amorphous phase. Above ~70 K, $C_2H_6$ undergoes rapid sublimation in our vacuum system, and so all data reported here are for lower temperatures. Band shapes, intensities, widths, and positions are relatively independent of temperature for each of the ethane phases, but differ significantly among the three phases studied, as seen in Fig. 1 of Hudson et al. (2009).

The situation surrounding the IR spectra of ethylene ices is much less clear than for the case of ethane. Jacox (1962) published $C_2H_4$ spectra at 4 K and 53 K that showed distinct differences in peak positions, splittings, and line widths. Concurrently, Wieder and Dows (1962) reported ethylene spectra at 63 K that matched the upper-temperature result of Jacox. Rytter and Gruen (1979) subsequently reexamined the spectra of ethylene ices at 15 - 85 K and concluded that Jacox's two spectra represented different crystalline phases, as she suggested, and that an irreversible phase change occurred between them, probably near 45 K. Still later, Zhao et al. (1988) questioned this interpretation, but as they showed only a single spectrum, which matched Jacox's 4-K result, detailed comparisons with earlier work are difficult, although by analogy with ethane, two crystalline phases of $C_2H_4$ seem reasonable. Each of the earlier papers included considerable information on band assignments in terms of ethylene's known crystal structure near 85 K or higher (Van Nes and Vos, 1978 and references therein), but none included measurements of an index of refraction from which optical constants could be calculated. It also appears that the IR spectra of *amorphous* ethylene, as with the case of amorphous ethane, have not been presented in any detail, if at all.

The published IR optical constants for $C_2H_6$ and $C_2H_4$ are limited to two papers. The optical constants reported for $C_2H_6$ are for an ice that was grown at ~15 K, annealed at an unspecified temperature, and then cooled to 30 K (Pearl et al., 1991). About twenty *k* values, but no $n$ values, were published for the 6000 - 5000 cm$^{-1}$ region along with about fifty ($n$, $k$) pairs from 5000 to 450 cm$^{-1}$. For ethylene, Zhao et al. (1988) showed the spectrum of a crystalline sample of unstated thermal history, recorded at 55 K from 4000 to 450 cm$^{-1}$. The computed $n$ and $k$ values were shown only in graphs, not listed in tables. In short, the published optical constants of $C_2H_6$ and $C_2H_4$ consist of one set of spectra and computations for each ice at a single temperature and only one phase. Quirico and Schmitt (1997) have published an extensive set of near-IR absorption coefficients, but not optical constants.

In this paper we present our recent measurements of near- and mid-infrared optical constants for multiple phases of ethane ($C_2H_6$) and ethylene ($C_2H_4$) from roughly 10 to 60 K. As in our recent work on $C_2H_2$, we also report new measurements of the index of refraction of each ice at 670 nm. Comparisons are made to earlier results where possible and electronic versions of our new data are made available. One of our primary goals for this paper is to link and compare the results already in print for the near-IR region, of importance to TNO observers, with data for the mid-IR spectral region, which has received far more attention by laboratory spectroscopists. Having data in both regions was particularly useful for identifying hydrocarbon ice phases, since relatively few spectra have been published for hydrocarbon ices in the near-IR as compared to the mid-IR region.

## 2. Laboratory procedures and calculations

The equipment used and methods followed in the present study were similar to those of Hudson et al. (2014). Ethane was purchased from Air Products (CP grade) and ethylene from Matheson (research grade, 99.99%), and both reagents were used as received. As before, a double-laser system was used to determine the index of refraction of ices at 670 nm, and these $n_{670}$ values then were used to measure the thickness of each ice and to calculate optical constants. Growth rates for ices were on the order of 1 μm hr$^{-1}$, with ices for mid-IR measurements restricted to a thickness of 0.5 - 1 μm or less, but ices for near-IR spectra, with their weaker features, being up to five times larger. Warming and cooling rates were about 3 K min$^{-1}$. A spectral resolution of 1 cm$^{-1}$ was used, ices were formed on a KBr substrate, and transmission spectra were generated from 100 scans with the IR beam of the spectrometer normal to the surface of the substrate. Checks were carried out to ensure that the IR features of interest were neither distorted by the spectral resolution used nor saturated by the ice thickness selected. As before, an iterative Kramers-Kronig method was used to calculate optical constants, with the calculated spectrum matching the observed spectrum within 10$^{-5}$ and reproducing reflection, absorption, and interference effects (Moore et al., 2010; Hudson et al., 2014).

Errors in our calculations derive mainly from the measurement of ice thickness, which in turn depends on values for $n_{670}$ and the counting of interference fringes. Our propagation-of-errors analysis (Taylor, 1997; Garland et al., 2003) gave an uncertainty in $n_{670}$ of ±0.015, or about ~1% of $n_{670}$ itself, and errors in the corresponding band strengths ($A$) and absorption coefficients ($\alpha$) near 6% for thinner ices and 0.6% for thicker ones (Hudson et al., 2014).

The focus of this manuscript is the stronger fundamental bands of ethane and ethylene in the mid-IR region, from 4000 to 400 cm$^{-1}$, and the near-IR features from 5000 to 4000 cm$^{-1}$.

Weaker features also were observed and recorded, mainly overtone and combination bands, but not examined in detail.  Some artifacts and spurious results worth mentioning are (i) the weak, but sharp, background lines of atmospheric $H_2O$ (3900 - 3600, 1900 - 1300 $cm^{-1}$) and $CO_2$ (2400 - 2300 $cm^{-1}$), resulting from incomplete background subtraction, and (ii) the weak, broad lines from trace amounts of $H_2O$ and $CO_2$ (2340 $cm^{-1}$) in the starting materials, especially the faint features of $C_2H_4$-$H_2O$ complexes (Thompson et al., 2005) in the 3700 - 3600 $cm^{-1}$ region.  The noise level in the spectra corresponds to a value of about 5 $cm^{-1}$ in terms of an absorption coefficient ($\alpha$); values of $\alpha$ less than about 5 $cm^{-1}$, as calculated from our tables, probably are unreliable.  The maximum intensity of the atmospheric $H_2O$ lines already mentioned corresponds to about 20 $cm^{-1}$.  Wavenumber positions for ethane and ethylene are listed in our Tables 2 and 3 to 0.1 $cm^{-1}$, calculated and measured absorption coefficients are rounded to four significant figures, and calculated and measured band strengths are rounded to three significant figures, largely due to uncertainties in ice densities.

## 3. Results

We first present our $n_{670}$ measurements and then our IR spectra of ethane and ethylene ices, including the influence of temperature on the spectra.

### 3.1. Refractive indices for $C_2H_4$ and $C_2H_6$ ices at 670 nm

Reference indices of refraction were required for optical-constants calculations of solid ethane and ethylene.  As we are unaware of any such published values at the temperatures of interest, we measured $n_{670}$ for amorphous, metastable, and crystalline ethane and ethylene.  Table 1 lists the results, with each tabulated $n_{670}$ value being the average of three measurements.  No attempts have been made to analyze either the variations among phase or molecule or any changes that might be due to temperature.   For a recent paper in this journal using the same laboratory techniques we employed to measure $n_{670}$, see Romanescu et al. (2010).  The method is similar to that pioneered by Tempelmeyer et al. (1968).

### 3.2. Infrared spectra of $C_2H_6$ ices

Figure 1 shows near- and mid-IR spectra of the three phases of ethane studied, each sample being brought to a common temperature of 20 K and having a thickness of about 0.5 $\mu$m.  Again, we note that crystalline $C_2H_6$ can be made either by direct deposition near 65 K or by warming one of the other two phases, but the metastable phase can only be achieved by direct deposition at 30-50 K.  In Fig. 2 we show expansions of several key regions, including the near-IR features of special interest to TNO observers.  Spectra of all three phases agree qualitatively with our earlier work (Hudson et al. 2009) and with the amorphous-ethane spectrum of Boudin et al. (1998), the metastable-ethane spectrum of Wisnosky et al. (1983), and the crystalline-ethane spectrum of Pearl et al. (1991).  More-quantitative comparisons appear later in this paper.  Our new work has many advantages over previous studies, including the advantage of all measurements coming from one laboratory, the greater number of ice phases and temperatures examined, the higher spectral resolution employed, and the increased accuracy in measuring each ice's thickness, which raises the reliability of the optical constants and band strengths derived.

### 3.3. Infrared spectra of $C_2H_4$ ices

Figure 3 shows the near- and mid-IR spectra of ethylene in three different phases, again all at 20 K. The bottom trace has absorptions that are broad and relatively featureless compared to spectra at the higher temperatures, suggesting that the former are for an amorphous $C_2H_4$ ice. By analogy with ethane, we interpret the middle spectrum as that of a low-temperature, metastable crystalline form of ethylene that cannot be achieved by warming but only by direct deposition. Figure 4 shows expansions of several ethylene bands for all three phases at 20 K. Warming an amorphous $C_2H_4$ sample gave distinct spectral changes near 35 K and again near 50 K, as seen in Fig. 5. The first change produced a mixture of the two higher-temperature crystalline forms, while the second gave an almost-completely high-temperature crystalline phase, each change being irreversible. The spectral changes corresponding to warmings below ~50 K were somewhat subtle for most of ethylene's IR features, as seen in the 4500-cm$^{-1}$ region of Fig. 5, but the small 822-cm$^{-1}$ band shown was found to be a very sensitive indicator of phase changes. All things considered, the spectral features we observed for metastable and crystalline ethylene were similar to those reported in the literature.

Of the six hydrocarbon ice phases examined in this paper, amorphous ethylene was the most difficult to prepare. Deposition rates of ~1 μm hr$^{-1}$ or lower and temperatures below ~20 K were needed for reproducible amorphous $C_2H_4$ ices. Samples grown above ~20 K at that same rate were at least partially crystalline, and ethylene ices grown faster than a few μm hr$^{-1}$ were partially or fully crystalline regardless of the temperature of formation. The possible influence of deposition speed in determining an ice's phase has been noted before (Jacox, 1962; Rytter and Gruen, 1979), and we believe that the present work conclusively demonstrates the influence of condensation rate for the case of solid ethylene.

Metastable ethylene was another case in which deposition rate was important. The only published spectra for this ice phase of which we are aware are those of Jacox (1962), although others have described their own results. We found that the key to reproducing the older spectra was to deposit $C_2H_4$ at rates far greater than what we employed in all other experiments, specifically about 60 μm hr$^{-1}$ as opposed to our usual ~1 μm hr$^{-1}$. Related to this, we note that for the ethylene experiments of Rytter and Gruen (1979), the condensation rates used were 75 - 750 μm hr$^{-1}$, and only metastable and crystalline $C_2H_4$ resulted. See also Hudson et al. (2014) for an example of how deposition differences can explain the variation between an older study of acetylene (Boudin et al., 1998) and our more-recent one.

Although some spectra of $C_2H_4$ ices have been published, those in Figs. 3 - 5, like similar ones for ethane, have several advantages over the older work. These include the advantage of a more-careful consideration of temperature, a direct measurement of ice thickness based on a recently-measured $n_{670}$ value, spectra having been recorded at more temperatures and ice phases than in earlier studies, and an explicit consideration of the effects of warming and cooling the samples. Our figures and the corresponding tables (*vide infra*) include data for only six temperatures, as the spectral variations with temperature were small for each phase compared to variations between phases. Other than at phase transitions, warming or cooling of ices gave only relatively small shifts in peak positions, band intensities, and line shapes.

Finally, we note that warming an amorphous or metastable ice, either $C_2H_4$ or $C_2H_6$, to 60 K, the highest temperature studied, always resulted in crystallization, but that direct deposition of a sample at 60 K always gave slightly sharper IR features for the crystalline solid. Similar observations with other molecules have been made by us (Moore et al, 2010) and others (Smith et al., 1994).

## 4. Discussion

*4.1. Optical constants, spectra, and spectral intensities of $C_2H_6$ and $C_2H_4$ ices*

The spectra just described and the values of $n_{670}$ measured were combined with a Kramers-Kronig routine (Moore et al., 2010) to calculate infrared optical constants, *n* and *k*, for (i) amorphous ices warmed from ~12 K, (ii) crystalline ices cooled from ~65 K, and (iii) metastable-phase ices made at an intermediate temperature and cooled, then warmed. As an example of these results, Fig. 6 shows *n*, *k*, and the measured transmission spectra for crystalline $C_2H_6$ and crystalline $C_2H_4$. Figures 7 and 8 present values of *n* and *k* at 20 K for each of the three phases of ethane and ethylene, respectively, in the same spectral regions given in Figs. 2 and 4. These optical constants and the spectra of Fig. 6 are posted at our lab's web site[1] along with similar results at a range of temperatures for each of the three ethane and ethylene phases studied. The *n* and *k* values posted will be useful for computing TNO spectra, and the spectral files should facilitate interlaboratory comparisons and the checking of computer routines for calculating optical constants.

Beyond these applications of our *n* and *k* values, our results also can be used to determine at least three measures of spectral band strength (Hudson et al., 2014). First, an absorption coefficient ($\alpha$) at any wavenumber ($\tilde{v}$), such as for an IR absorption peak, can be calculated from our *k* values using $\alpha(\tilde{v}) = 4\pi\tilde{v}k(\tilde{v})$. Second, any set of computed $\alpha(\tilde{v})$ values can be integrated over an IR band of interest as

$$\int \alpha(\tilde{v})\,d\tilde{v} = \int 4\pi\tilde{v}k(\tilde{v})\,d\tilde{v} \qquad (1)$$

to give an integrated intensity. Finally, if the number density ($\rho_N$, molecules cm$^{-3}$) of absorbers in an ice is either known or can be estimated then

$$A = \frac{1}{\rho_N} \int_{band} \alpha(\tilde{v})\,d\tilde{v} \qquad (2)$$

gives a band strength, here denoted *A* (and typically in units of cm molecule$^{-1}$). From *A* it is possible to use astronomical spectra to derive molecular abundances, as is the common practice. For a recent example, see Yamagashi et al. (2011).

---

[1] http://science.gsfc.nasa.gov/691/cosmicice/constants.html

These spectral intensities determined from optical constants sometimes are called *absolute* absorption coefficients ($\alpha$) and absolute band strengths ($A$) in order to distinguish them from *apparent* absorption coefficients ($\alpha'$) and apparent band strengths ($A'$), which are calculated directly from laboratory spectra. It has been known for over 50 years (Maeda and Schatz, 1961) that $\alpha'$ and $A'$, although easy to evaluate from lab spectra, can include intensity losses due to reflection and interference effects, which can be significant for thin ices. The differences can be difficult to predict *a priori*, and so $\alpha$, integrated $\alpha$ values, and $A$ should be used when available. A corollary is that $k$ alone will not necessarily give an accurate calculation of the IR spectrum of a thin ice. For comparisons of $\alpha$ and $A$ of crystalline acetylene with $\alpha'$ and $A'$, respectively, see Hudson et al. (2014).

Tables 2 and 3 summarize our spectral-intensity calculations for ethane and ethylene. The first column of each table lists selected spectral peaks. Graphs of each peak's height (absorbance) as a function of ice thickness gave Beer's-law lines whose slopes are $\alpha' / \ln(10)$. The second columns in Tables 2 and 3 list $\alpha'$ values. Following calculations of optical constants, plots of $k$ were examined for peak positions, and these are listed in the third columns of Tables 2 and 3, followed by columns with calculations of $\alpha$ from $\alpha(\tilde{\nu}) = 4\pi\tilde{\nu}k(\tilde{\nu})$. Finally, integrations of $\alpha$ and $\alpha'$ over the ranges listed in each table, and application of equations (1) and (2), gave the absolute and apparent band strengths, $A$ and $A'$, respectively. The differences, particularly for ethane, between $\alpha'$ and $\alpha$, and between $A$ and $A'$, vary considerably and are difficult to predict in advance, emphasizing the need for optical-constants calculations for accurate intensity values.

*4.2. Comparisons to previous work*

Comparisons of our results to published work are considerably hindered by a lack of experimental details in most of the earlier work. For example, the $A'$ values of our Table 3 are comparable to two values published by Dows (1966), but the density assumed by the latter is unknown, precluding a meaningful quantitative comparison. The only published mid-IR optical constants of which we are aware for $C_2H_6$ are those of Pearl et al. (1991) for crystalline ethane at 30 K. Our *n*-and-*k* results for amorphous ethane ices crystallized by warming from ~12 K, and then recooling to 30 K, are essentially the same as those of Pearl et al. (1991), but crystalline ices grown at 60 K and then cooled to 30 K gave intensities that were up to a factor of two greater. Relevant details concerning thickness measurements are missing from Pearl et al. (1991), again preventing a more-careful investigation. Perhaps the best comparison is Fig. 9, which shows our crystalline-ethane absorption coefficients ($\alpha$) in the near-IR region with points added for the results reported by Pearl et al. (1991) and Quirico and Schmitt (1997). For the most part, the results of all three research groups agree.

Turning to $C_2H_4$, to our knowledge the only IR optical constants available are those of Zhao et al. (1988). Unfortunately, a quantitative comparison of our results with theirs is impossible for several reasons: (1) Their paper lacks details on how ice thicknesses were measured. The usual measurement techniques, as in the present paper, involve interference fringes and a reference index of refraction to convert a fringe count into an ice thickness. However, no such value was provided by Zhao et al. (1988), making it impossible to compare accurately their ice thicknesses and integrated intensities to ours. (2) The calculated optical constants of Zhao et al. (1988) were not tabulated, but rather were presented in figures with small scales (no enlargements), showing few details of IR bands. (3) The phase of the ethylene ice studied by Zhao et al. (1988) is uncertain. The authors state that their sample was at 55 K, which means that it was in the *high*-temperature crystalline phase, but their Table 2 gives three components for ice's $\nu_{10}$ band (~822 cm$^{-1}$), which means that their sample was *not* in the high-temperature crystalline phase (see our Fig. 4). Adding to the confusion, the only $C_2H_4$ spectrum shown (their Fig. 1) appears to have only *two* components for the $\nu_{10}$ band, not the three they tabulated, and to be a good match for the high-temperature phase of $C_2H_4$ (Jacox, 1962; Wieder and Dows, 1962; our Fig. 4). These considerations make it impossible to know which of our *n* and *k* data sets should be compared to the results of Zhao et al. (1988) on ethylene.

Two other ethylene comparisons are possible. Wieder and Dows (1962) reported band-strength measurements for crystalline ethylene at 63 K. In general, their values agree with ours, although again more experimental details about their work (e.g., ice density) would make the comparison more quantitative. The final ethylene comparison is shown in Fig. 10 with our absorption coefficients plotted alongside those of Quirico and Schmitt (1997). Once more, the agreement is good.

*4.3. Additional comments*

Of the six hydrocarbon ice phases covered in this paper, only five are likely to be important for astronomical objects. The formation of metastable ethylene ice required a much faster deposition rate than that for the other phases we examined, so fast that it would appear to be astronomically unrealistic.

Another point worth noting is that the near-IR features for ethane, and also for ethylene, do not change strongly with temperature and phase, aside from some sharpening for the crystalline ices. This will make the correlation of specific astronomical spectra with temperature difficult with $C_2H_6$ and $C_2H_4$. Spectral features of other molecules may well be much better indicators of temperature.

In this paper we have focused on the IR optical constants and band strengths of ethane and ethylene ices, results of astrochemical interest. However, the electronic versions of our data permit the examination of other spectral results, such as line widths and the relatively-small band-strength variations seen with temperature changes. Our archived spectra, particularly in the case of ethylene, also possess small features corresponding to forbidden spectral transitions, which we have not explored.

## 5. Conclusions

We have measured near- and mid-IR spectra for three phases each of ethane and ethylene ices from ~5000 to 500 cm$^{-1}$ (1.43 to 20 µm) at multiple temperatures, and have calculated the corresponding optical constants *n* and *k*. All IR spectra and resulting sets of $n(\tilde{\nu})$ and $k(\tilde{\nu})$ are available in electronic form on our web site (http://science.gsfc.nasa.gov/691/cosmicice/ constants.html). We also have tabulated absolute and apparent band intensities, presenting differences in the two that highlight the need for optical-constants calculations for precision spectral fitting, analysis, and interpretation.

One goal of this paper has been to link the earlier mid-IR laboratory spectroscopic measurements on ethane and ethylene with the much less common near-IR studies (e.g., Quirico and Schmitt, 1997) needed for planetary work. Our figures and tables show that this goal has been met. We emphasize that much of the older data is unsuitable for quantitative studies with either planetary or interstellar observations due to the uncertainties in the conditions under which the lab data were recorded.

**Acknowledgments**

NASA funding through the Outer Planets Research and Cassini Data Analysis programs is acknowledged. RLH and PAG received partial support from the NASA Astrobiology Institute through the Goddard Center for Astrobiology. Karen Smith (NPP, NASA-GSFC) is thanked for laboratory assistance.

Table 1.

$n_{670}$ for $C_2H_6$ and $C_2H_4$ Ices

| Molecule | Phase | $n_{670}$ | Measurement Temperature (K) |
|---|---|---|---|
| $C_2H_6$ | amorphous | 1.34 | 12 |
| $C_2H_6$ | metastable[a] | 1.41 | 47 |
| $C_2H_6$ | crystalline[b] | 1.44 | 60 |
| $C_2H_4$ | amorphous | 1.35 | 11 |
| $C_2H_4$ | metastable[a] | 1.35 | 30 |
| $C_2H_4$ | crystalline[b] | 1.45 | 45 |

[a] metastable, low-temperature crystalline phase
[b] high-temperature crystalline phase

Table 2

Positions and Strengths for Selected $C_2H_6$ Features

| $\tilde{v}$ (cm$^{-1}$)[a] | $\alpha'$ (cm$^{-1}$)[b] | $\tilde{v}$ (cm$^{-1}$)[c] | $\alpha$ (cm$^{-1}$)[d] | Integration range (cm$^{-1}$) | $A'$ (10$^{-18}$ cm molec$^{-1}$)[e] | $A$ (10$^{-18}$ cm molec$^{-1}$)[e] |
|---|---|---|---|---|---|---|
| \multicolumn{7}{c}{Amorphous $C_2H_6$ at 20 K} | | | | | | |
| 4322.1 | 307.7 | 4321.9 | 301.9 | 4339-4293 | 0.218 | 0.173 |
| 4161.3 | 260.2 | 4161.6 | 255.5 | 4222-4140 | 0.238 | 0.209 |
| 4064.6 | 219.3 | 4064.7 | 222.4 | 4087-4047 | 0.155 | 0.125 |
| 2972.3 | 22720 | 2972.0 | 21990 | 3020-2892 | 22.0 | 20.0 |
| 2880.2 | 8672 | 2880.3 | 7899 | 2892-2868 | 3.81 | 3.51 |
| 1462.4 | 2706 | 1462.6 | 2318 | 1514-1430 | 3.76 | 3.30 |
| 1368.9 | 1915 | 1368.9 | 1592 | 1379-1361 | 0.618 | 0.530 |
| 817.1 | 3328 | 817.1 | 2826 | 835-810 | 1.99 | 1.69 |
| \multicolumn{7}{c}{Metastable $C_2H_6$ at 40 K} | | | | | | |
| 4321.9 | 631.0 | 4322.0 | 734.8 | 4337-4295 | 0.357 | 0.424 |
| 4159.9 | 367.3 | 4159.3 | 324.9 | 4200-4141 | 0.257 | 0.246 |
| 4065.1 | 538.3 | 4065.1 | 647.6 | 4085-4048 | 0.190 | 0.223 |
| 2971.4 | 54220 | 2971.7 | 60390 | 3020-2890 | 27.3 | 29.2 |
| 2880.0 | 3687 | 2880.1 | 7084 | 2895-2866 | 0.775 | 1.46 |
| 1463.0 | 9072 | 1463.2 | 9821 | 1500-1420 | 5.12 | 4.84 |
| 1452.9 | 12540 | 1453.1 | 14190 | | | |
| 1369.3 | 1332 | 1369.4 | 2353 | 1380-1355 | 0.178 | 0.293 |
| 821.2 | 11990 | 821.5 | 11750 | 835-800 | 3.84 | 3.50 |
| 814.9 | 14810 | 815.1 | 13170 | | | |

[a] taken directly from the observed spectra; [b] obtained from the slopes of Beer's Law graphs; [c] peaks taken from the calculated spectrum of $k$ values; [d] calculated from $k$ using $\alpha = 4\pi\tilde{v}k$. [e] $A'$ and $A$ are calculated from $A' = (\int \alpha' d\tilde{v} \times MW) / (\text{density} \times N_A)$ and $A = (\int \alpha d\tilde{v} \times MW) / (\text{density} \times N_A)$, respectively, where MW = molecular weight = 30.069 g mol$^{-1}$, $N_A$ = 6.022 × 10$^{23}$ molecules mol$^{-1}$, and density = 0.719 g cm$^{-3}$ (van Nes and Vos, 1978).

Table 2 (Continued)

Positions and Strengths for Selected $C_2H_6$ Features

| $\tilde{\nu}$ (cm$^{-1}$)[a] | $\alpha'$ (cm$^{-1}$)[b] | $\tilde{\nu}$ (cm$^{-1}$)[c] | $\alpha$ (cm$^{-1}$)[d] | Integration range (cm$^{-1}$) | $A'$ (10$^{-18}$ cm molec$^{-1}$)[e] | $A$ (10$^{-18}$ cm molec$^{-1}$)[e] |
|---|---|---|---|---|---|---|
| Crystalline $C_2H_6$ at 60 K | | | | | | |
| 4320.2 | 584.1 | 4320.3 | 531.2 | 4338-4295 | 0.383 | 0.372 |
| 4158.5 | 494.2 | 4158.6 | 475.9 | 4200-4141 | 0.333 | 0.410 |
| 4064.3 | 510.1 | 4064.3 | 483.7 | 4085-4048 | 0.202 | 0.207 |
| 2972.0 | 54300 | 2972.3 | 56210 | 3020-2892 | 30.2 | 31.4 |
| 2880.1 | 22830 | 2880.3 | 22950 | 2895-2866 | 4.75 | 5.10 |
| 1466.2 | 2068 | 1466.5 | 1809 | | | |
| 1456.8 | 8600 | 1456.9 | 9533 | 1500-1420 | 4.41 | 4.24 |
| 1450.4 | 3708 | 1450.6 | 3147 | | | |
| 1370.4 | 5212 | 1370.5 | 4851 | 1392-1348 | 1.39 | 1.32 |
| 1368.6 | 6141 | 1368.8 | 4907 | | | |
| 825.2 | 11700 | 825.7 | 11010 | | | |
| 816.6 | 9956 | 816.9 | 10070 | 835-800 | 3.46 | 3.16 |
| 813.8 | 6459 | 813.9 | 5200 | | | |

[a] taken directly from the observed spectra; [b] obtained from the slopes of Beer's Law graphs; [c] peaks taken from the calculated spectrum of $k$ values; [d] calculated from $k$ using $\alpha = 4\pi\tilde{\nu}k$. [e] $A'$ and $A$ are calculated from $A' = (\int \alpha' d\tilde{\nu} \times MW) / (density \times N_A)$ and $A = (\int \alpha d\tilde{\nu} \times MW) / (density \times N_A)$, respectively, where MW = molecular weight = 30.069 g mol$^{-1}$, $N_A = 6.022 \times 10^{23}$ molecules mol$^{-1}$, and density = 0.719 g cm$^{-3}$ (van Nes and Vos, 1978).

Table 3

Positions and Strengths for Selected $C_2H_4$ Features

| $\tilde{\nu}$ (cm$^{-1}$)[a] | $\alpha'$ (cm$^{-1}$)[b] | $\tilde{\nu}$ (cm$^{-1}$)[c] | $\alpha$ (cm$^{-1}$)[d] | Integration range (cm$^{-1}$) | $A'$ (10$^{-18}$ cm molec$^{-1}$)[e] | $A$ (10$^{-18}$ cm molec$^{-1}$)[e] |
|---|---|---|---|---|---|---|
| | | | Amorphous $C_2H_4$ at 12 K | | | |
| 4706.5 | 143.9 | 4709.0 | 142.9 | 4722-4693 | 0.103 | 0.100 |
| 4495.9 | 359.1 | 4498.0 | 350.1 | 4520-4480 | 0.277 | 0.265 |
| 4189.1 | 185.4 | 4191.0 | 181.6 | 4210-4175 | 0.104 | 0.0956 |
| 3089.6 | 2883 | 3092.0 | 2875 | 3105-3073 | 1.51 | 1.52 |
| 2974.3 | 2516 | 2976.0 | 2510 | 2983-2968 | 1.03 | 1.02 |
| 1434.3 | 3841 | 1436.0 | 3988 | 1452-1428 | 2.24 | 2.31 |
| 949.3 | 9197 | 950.7 | 8915 | 1030-918 | 12.8 | 12.6 |
| 821.3 | 171.8 | 822.5 | 167.8 | 835-813 | 0.106 | 0.104 |
| | | | Metastable $C_2H_4$ at 20 K | | | |
| 4703.5 | 349.0 | 4703.7 | 310.3 | 4722-4693 | 0.141 | 0.139 |
| 4495.0 | 871.5 | 4495.1 | 795.8 | 4520-4480 | 0.322 | 0.324 |
| 4186.8 | 410.3 | 4186.9 | 367.6 | 4210-4175 | 0.127 | 0.129 |
| 3088.3 | 5807 | 3088.3 | 5978 | 3105-3073 | 1.92 | 1.74 |
| 2973.5 | 5263 | 2973.5 | 4694 | 2983-2968 | 0.755 | 0.842 |
| 1440.3 | 4208 | 1440.4 | 4931 | | | |
| 1437.0 | 2787 | 1437.1 | 2511 | 1448-1426 | 2.08 | 2.41 |
| 1433.6 | 6959 | 1433.5 | 7459 | | | |
| 950.4 | 13180 | 950.7 | 13440 | 1030-918 | 15.2 | 17.1 |
| 940.5 | 13620 | 940.8 | 14230 | | | |
| 825.7 | 85.04 | 825.7 | 67.16 | 829-814 | 0.149 | 0.156 |
| 822.9 | 892.4 | 822.9 | 997.4 | | | |
| 820.1 | 411.2 | 820.0 | 340.4 | | | |

[a] taken directly from the observed spectra; [b] obtained from the slopes of Beer's Law graphs; [c] peaks taken from the calculated spectrum of $k$ values; [d] calculated from $k$ using $\alpha = 4\pi\tilde{\nu}k$. [e] $A'$ and $A$ are calculated from $A' = (\int \alpha' d\tilde{\nu} \times MW) / (density \times N_A)$ and $A = (\int \alpha d\tilde{\nu} \times MW) / (density \times N_A)$, respectively, where MW = molecular weight = 28.053 g mol$^{-1}$, $N_A$ = 6.022 × 10$^{23}$ molecules mol$^{-1}$, and density = 0.75 g / cm$^{-3}$ (van Nes, 1978).

Table 3 (continued)

Positions and Strengths for Selected $C_2H_4$ Features

| $\tilde{\nu}$ (cm$^{-1}$)[a] | $\alpha'$ (cm$^{-1}$)[b] | $\tilde{\nu}$ (cm$^{-1}$)[c] | $\alpha$ (cm$^{-1}$)[d] | Integration range (cm$^{-1}$) | $A'$ (10$^{-18}$ cm molec$^{-1}$)[e] | $A$ (10$^{-18}$ cm molec$^{-1}$)[e] |
|---|---|---|---|---|---|---|
| Crystalline $C_2H_4$ at 60 K | | | | | | |
| 4704.7 | 368.8 | 4704.7 | 349.2 | 4722-4693 | 0.153 | 0.146 |
| 4497.4 | 1104 | 4497.3 | 1023 | 4525-4480 | 0.338 | 0.314 |
| 4188.9 | 522.4 | 4189.9 | 487.1 | 4210-4175 | 0.144 | 0.135 |
| 3089.3 | 5960 | 3089.1 | 5960 | 3105-3073 | 1.91 | 1.71 |
| 2974.3 | 5837 | 2974.1 | 5837 | 2982-2968 | 0.937 | 0.942 |
| 1440.7 | 9645 | 1440.8 | 9645 | 1452-1428 | 2.97 | 2.91 |
| 1436.5 | 9746 | 1436.5 | 9746 | | | |
| 949.2 | 11890 | 949.2 | 11990 | 1030-918 | 19.4 | 19.3 |
| 943.2 | 12660 | 943.2 | 11720 | | | |
| 825.6 | 228.5 | 825.7 | 188.8 | 829-814 | 0.0825 | 0.0884 |
| 819.8 | 919.4 | 819.8 | 731.8 | | | |

[a] taken directly from the observed spectra; [b] obtained from the slopes of Beer's Law graphs; [c] peaks taken from the calculated spectrum of $k$ values; [d] calculated from $k$ using $\alpha = 4 \pi \tilde{\nu} k$. [e] $A'$ and $A$ are calculated from $A' = (\int \alpha' \, d\tilde{\nu} \times MW) / (\text{density} \times N_A)$ and $A = (\int \alpha \, d\tilde{\nu} \times MW) / (\text{density} \times N_A)$, respectively, where MW = molecular weight = 28.053 g mol$^{-1}$, $N_A$ = 6.022 × 10$^{23}$ molecules mol$^{-1}$, and density = 0.75 g / cm$^{-3}$ (van Nes, 1978).

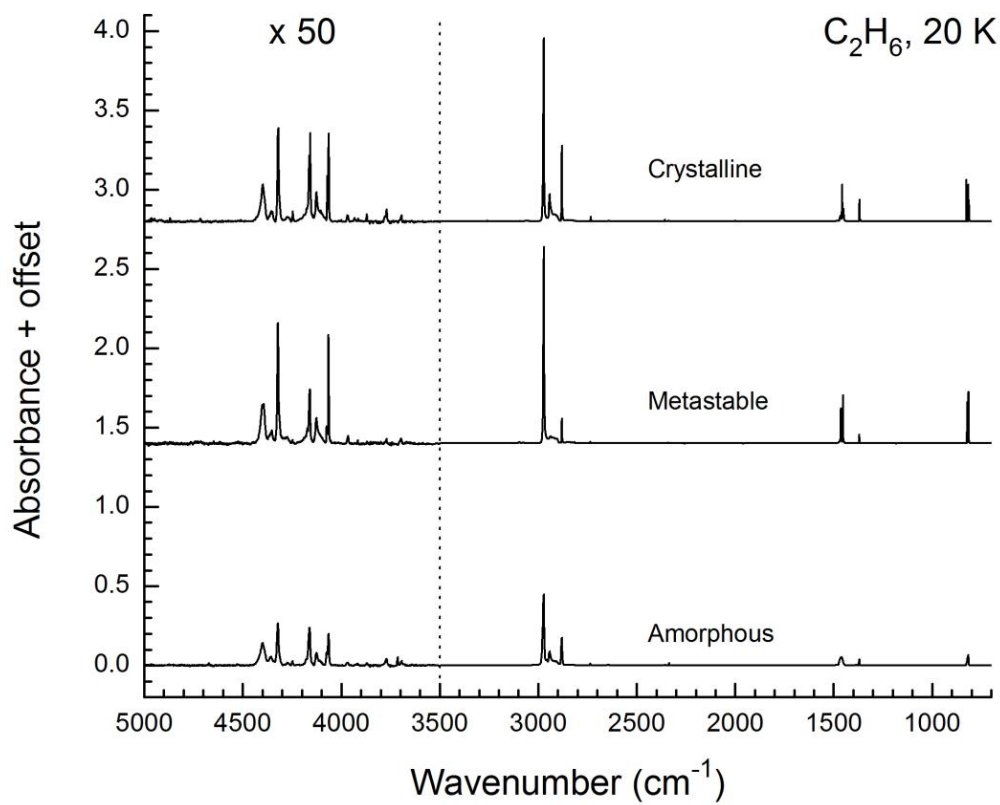

**Fig. 1.** Near- and mid-IR transmission spectra for three phases of $C_2H_6$ at 20 K. The ice thickness in each case was about 0.5 μm. The crystalline ice (top) was created at 60 K and cooled to 20 K. The metastable $C_2H_6$ ice (middle) was created at 45 K and cooled to 20 K. The amorphous ethane ice (bottom) was created at 12 K and heated to 20 K.

Hudson, R. L., et al., 2014, *Icarus* **243**, 148-157. DOI: 10.1016/j.icarus.2014.09.001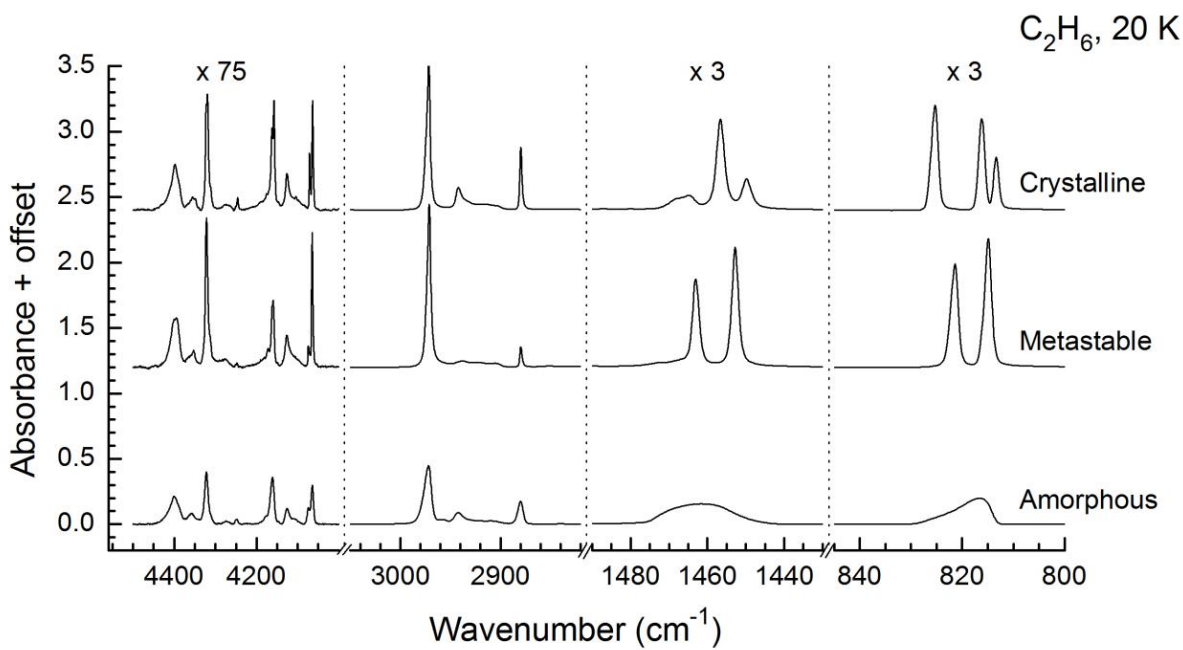

**Fig. 2.** Expansion of selected spectral regions of $C_2H_6$ ices showing details for each solid phase. The ice thickness in each case was about 0.5 μm. The crystalline ice (top) was created at 60 K and cooled to 20 K. The metastable $C_2H_6$ ice (middle) was created at 45 K and cooled to 20 K. The amorphous ethane ice (bottom) was created at 12 K and heated to 20 K.

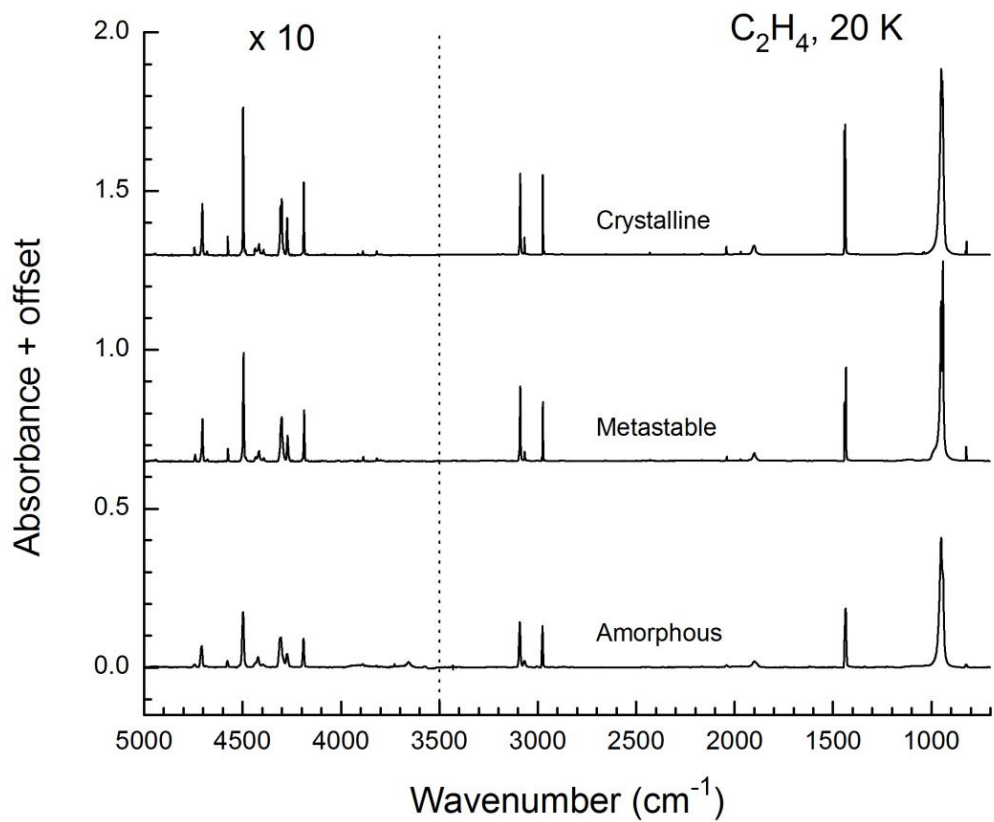

**Fig. 3.** Near- and mid-IR transmission spectra for three phases of $C_2H_4$ each at 20 K. The ice thickness in each case was about 1 μm. The crystalline ice (top) was created by a slow deposit (~1 μm hr$^{-1}$) at 60 K and then cooled to 20 K. The metastable $C_2H_4$ ice (middle) was created by a fast deposit (~60 μm hr$^{-1}$) at 20 K. The amorphous ethylene ice (bottom) was created by a slow deposit (~1 μm hr$^{-1}$) at 12 K and then heated to 20 K.

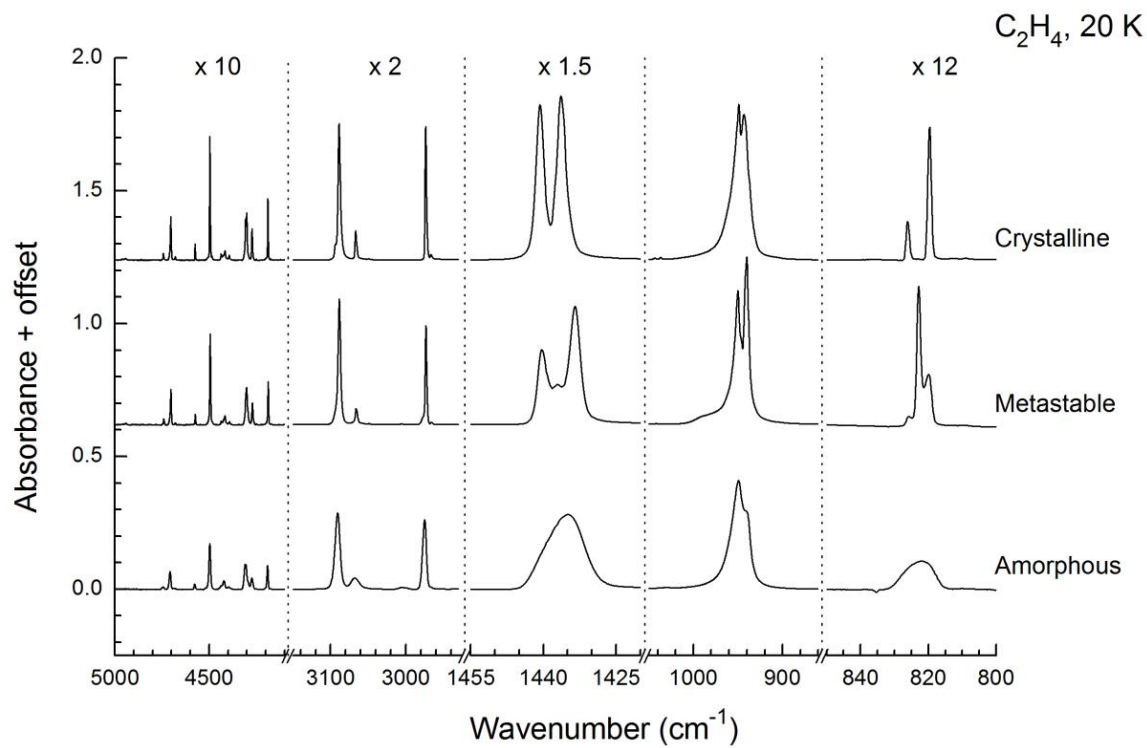

**Fig. 4.** Expansion of selected spectral regions of $C_2H_4$ ices showing details for each solid phase. The ice thickness in each case was about 1 μm. The crystalline ice (top) was created by a slow deposit (~1 μm hr$^{-1}$) at 60 K and then cooled to 20 K. The metastable $C_2H_4$ ice (middle) was created by a fast deposit (~60 μm hr$^{-1}$) at 20 K. The amorphous ethylene ice (bottom) was created by a slow deposit (~1 μm hr$^{-1}$) at 12 K and then heated to 20 K.

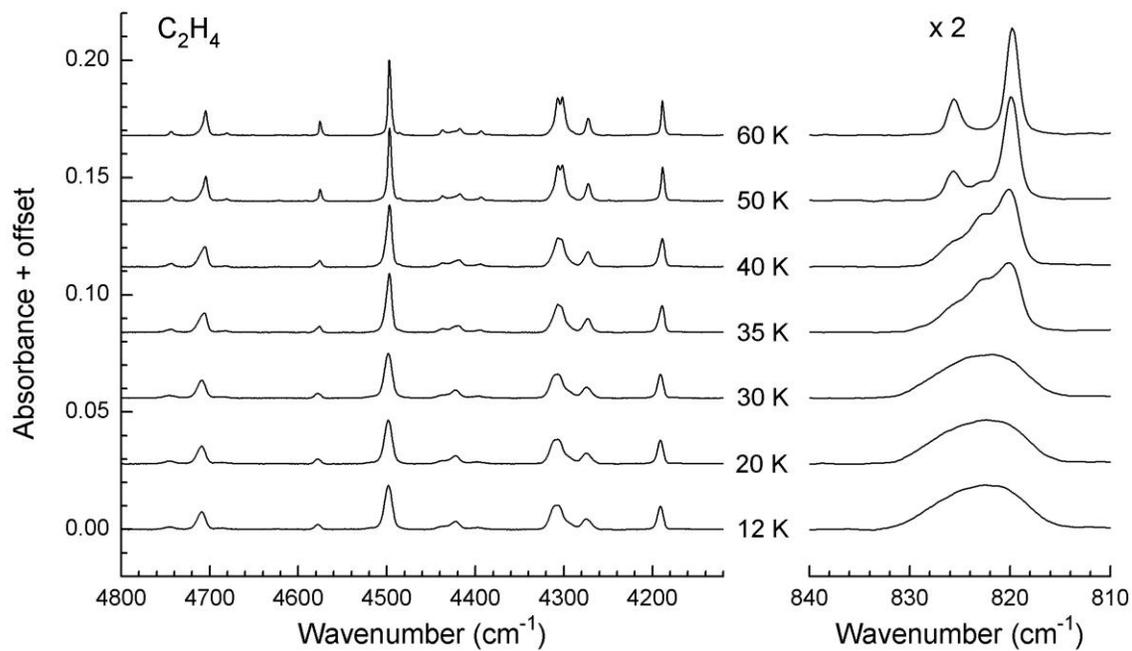

**Fig. 5.** IR spectra of a 1-μm $C_2H_4$ ice grown at 12 K and warmed to the temperatures indicated.

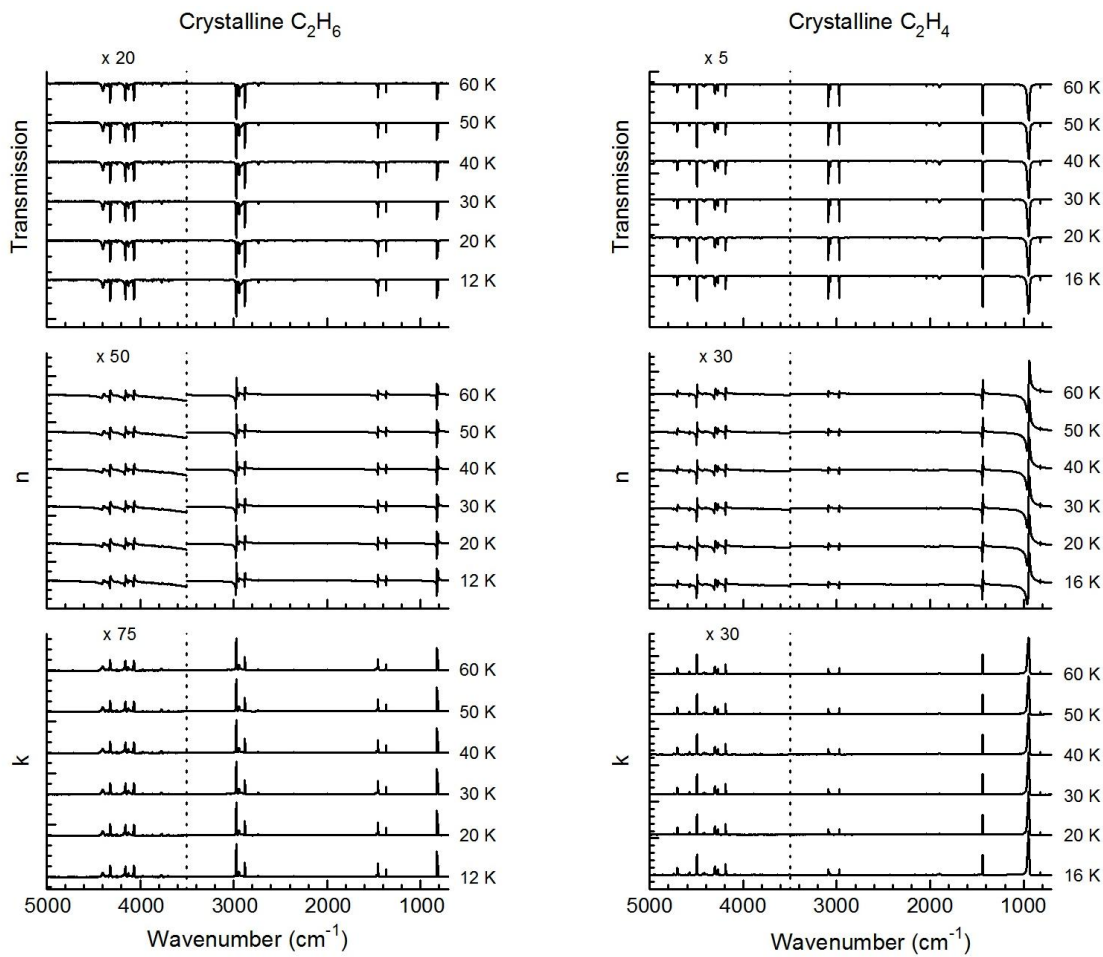

**Fig. 6.** IR transmission spectra and optical constants of crystalline $C_2H_6$ (left) and crystalline $C_2H_4$ (right) grown at 60 K and cooled to the temperatures indicated. Traces in each panel are stacked for clarity. The $C_2H_6$ sample's thickness was about 0.6 μm. For $C_2H_4$, the thickness of the ice used for the near-IR (5000 - 3500 cm$^{-1}$) was about 3 μm, while the thickness for the mid-IR (3500 - 700 cm$^{-1}$) sample was about 1 μm.

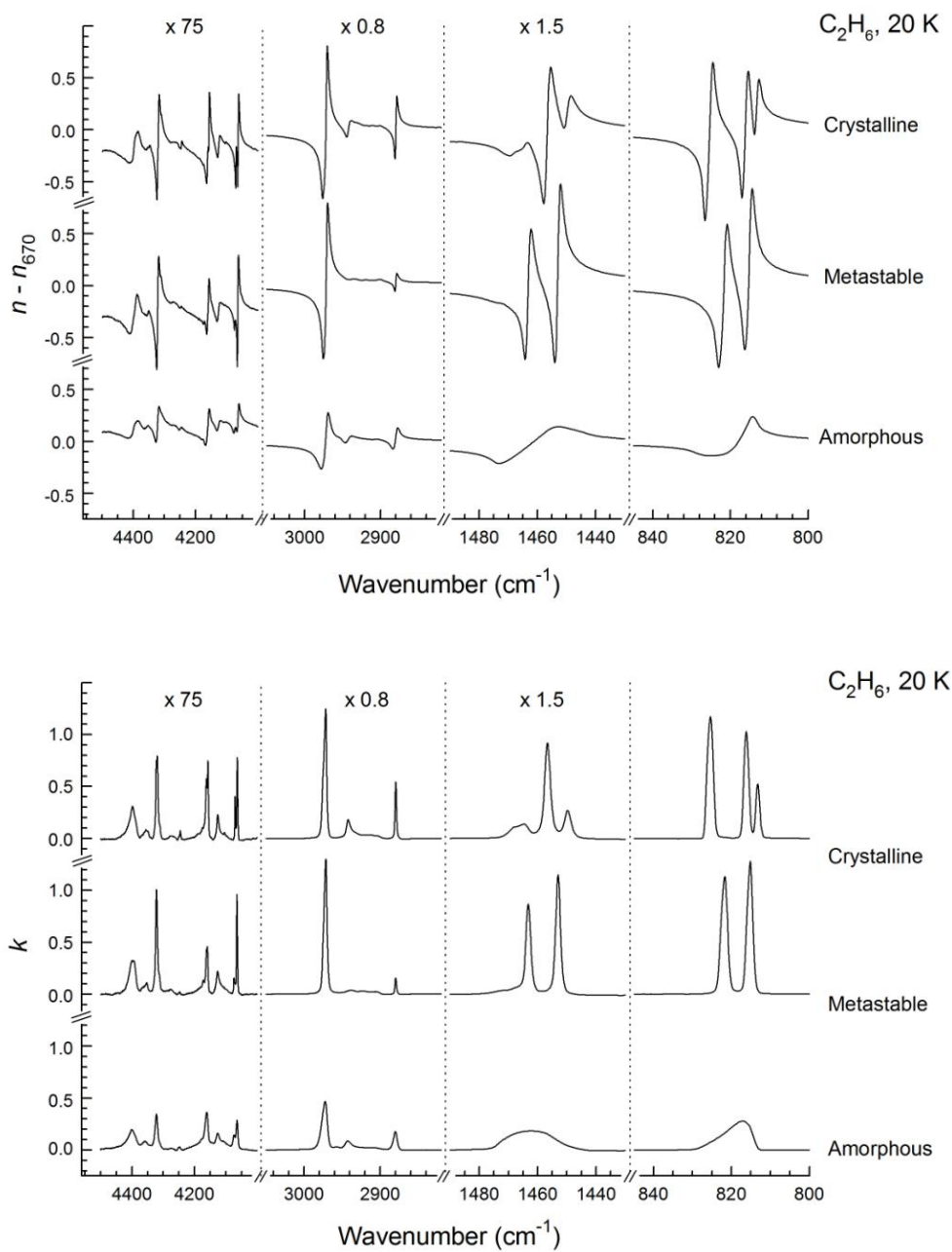

**Figure 7.** Calculated values of *n* and *k* for the three crystalline phases of ethane at 20 K. In the top panel, the phase-specific value of $n_{670}$ from Table 1 has been subtracted from *n* before scaling by the factor shown. Because they would otherwise fall outside the given plotting range, *n* values for the crystalline and amorphous phases in the 4500-4000 cm$^{-1}$ region were shifted by an additional +0.02 before scaling.

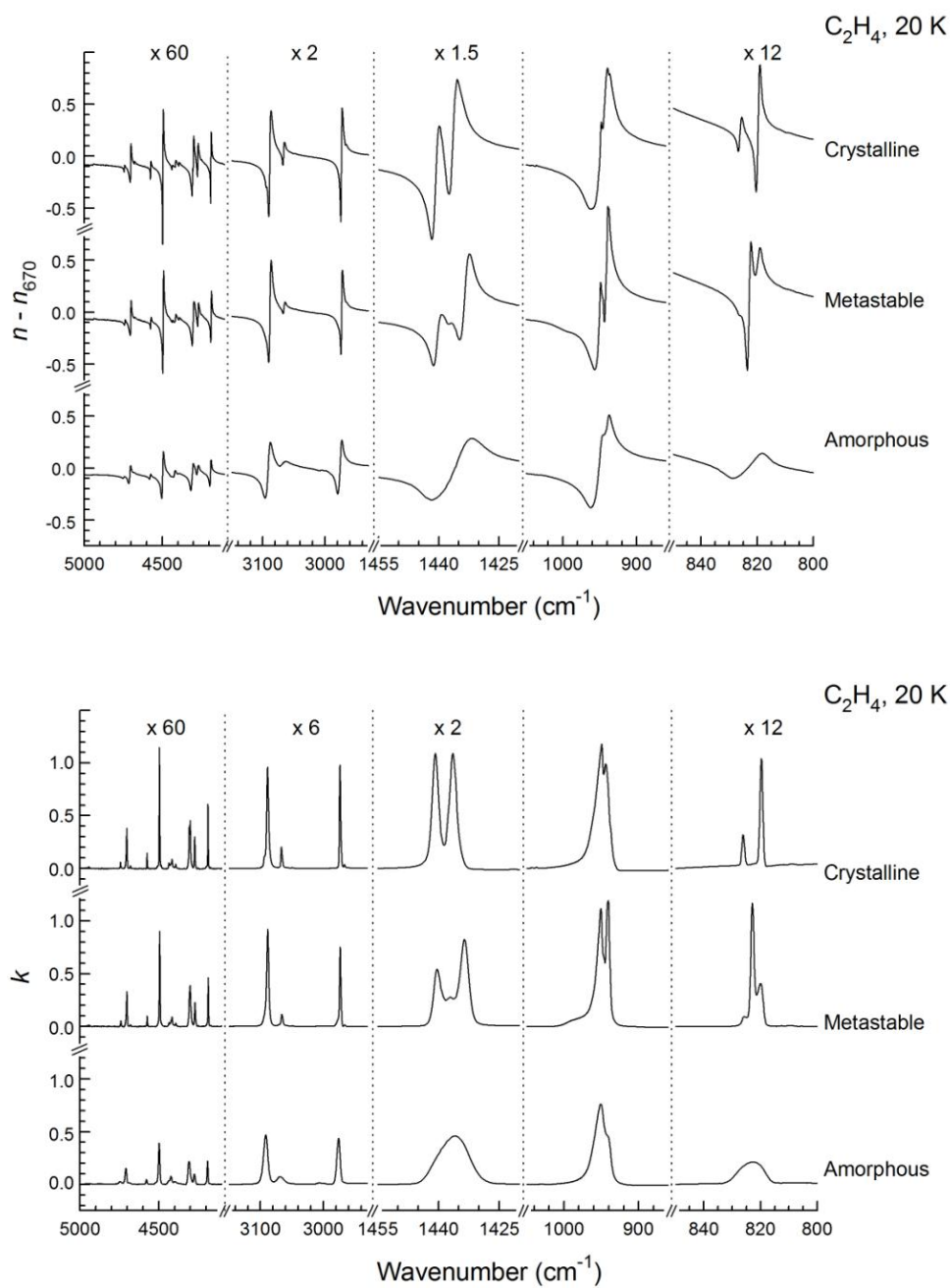

**Figure 8.** Calculated values of *n* and *k* for the three crystalline phases of ethylene at 20 K. In the top panel, the phase-specific values of $n_{670}$ from Table 1 have been subtracted from *n* before scaling by the factor shown. Because they would otherwise fall outside the given plotting range, *n* values in the 850-800 cm$^{-1}$ region for all three phases were shifted by an additional -0.05 before scaling.

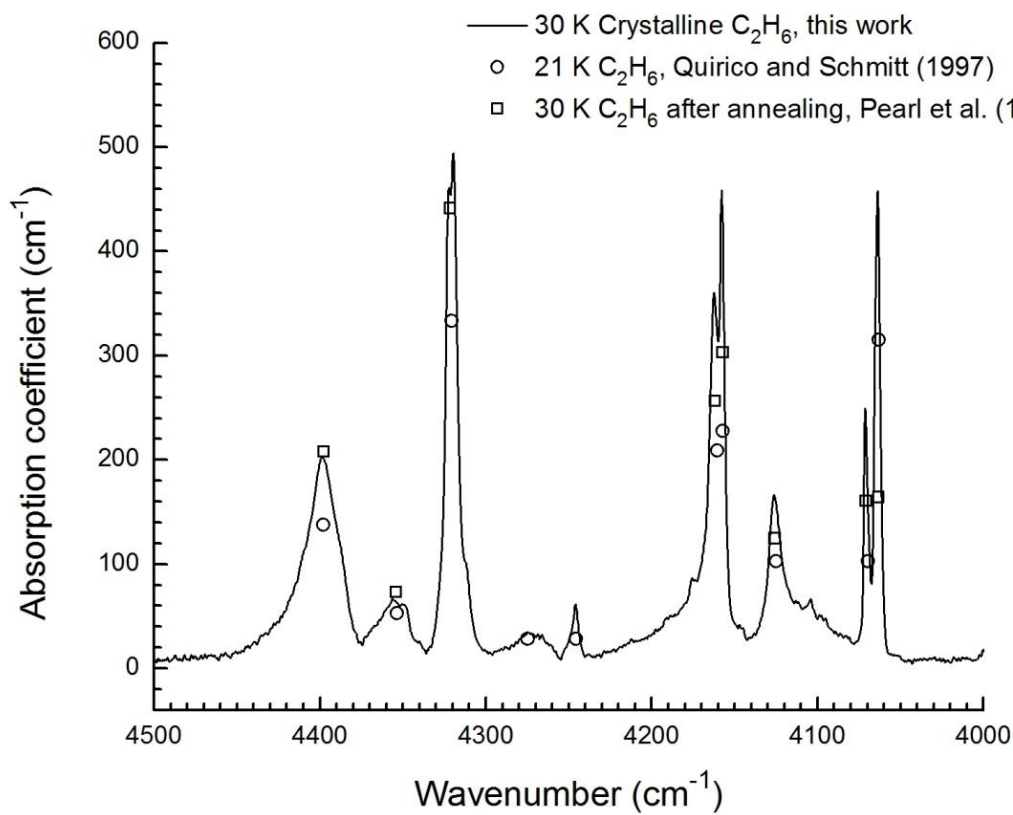

**Fig. 9.** Comparison of the present work (solid line) on $C_2H_6$ with the absorption coefficients of Pearl et al. (1991) and Quirico and Schmitt (1997).

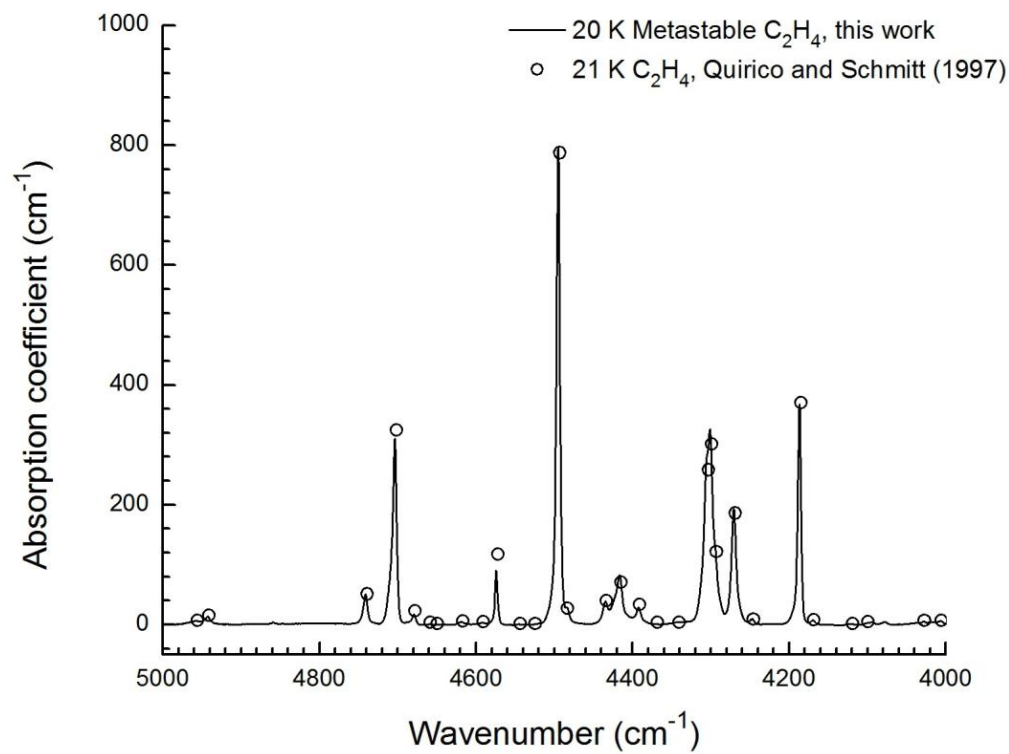

**Fig. 10.** Comparison of the present work (solid line) on $C_2H_4$ with the absorption coefficients of Quirico and Schmitt (1997).